\documentclass[12pt]{iopart}
 \usepackage{graphicx, iopams}
\usepackage{bm}
\usepackage[T1]{fontenc}
\usepackage[cp1250]{inputenc}

\newcommand{\nn}{\nonumber}
\newcommand{\AB}{\allowbreak}

\newcommand{\ali}[2]{\mathop{\mathfrak{#1}(#2)}\nolimits}

\newcommand{\ADA}[1]{\ifmmode \ad(#1) \else $\ad(#1)$\fi}
\newcommand{\LI}[2]{\ifmmode#2_1,\AB\,\ldots,\,\AB #2_{#1}%
\else$ #2_1,\AB\,\ldots,\,\AB#2_{#1}$\fi}
\newcommand{\bea}{\begin{eqnarray}}
\newcommand{\eea}{\end{eqnarray}}
\newcommand{\ba}{\begin{array}}
\newcommand{\ea}{\end{array}}
\newcommand{\iind}{$2^\mathrm{nd}~$}

\newcommand{\iiird}{$3^\mathrm{rd}~$}

\newcommand{\su}[1]{\ali{su}{#1}}
\newcommand{\sltwo}{\ifmmode \ali{sl}{2} \else $\ali{sl}{2}$\fi}

\long\def\comment#1{}

\newcommand{\bMA}[1]{\[\begin{array}{#1}}
\newcommand{\eMA}{\end{array}\]}

\newcommand{\C}{{{\mathbb C}}}

\newcommand{\I}{\mathbb{I}}

\newtheorem{corollary}{Corollary}

\newcommand{\NR}{{\mathbb{R}}}

\newcommand{\ivth}{$4^\mathrm{th}~$}

\def\be{\begin{equation}}
\def\ee{\end{equation}}
\def\bn{\begin{enumerate}}
\def\en{\end{enumerate}}

\def\R{\NR}

\def\zd{{z^\dagger}}
\def\Vd{{V^\dagger}}

\def\cpn{{\C P^{N-1}}}
\def\bp{{\bar{\partial}}}
\def\p{{\partial}}

\def\tr{{\mathrm{tr}}}
\def\var{{\mathrm{var}}}
\def\lam{\lambda}
\def\diag{\mathrm{diag}}

\newcommand{\qed}{\begin{flushright} $\square$
                  \end{flushright}
}
\newtheorem{proposition}{Proposition}
\begin{document}
\title[Soliton surfaces associated with sigma models]{Soliton surfaces associated with sigma models; differential and algebraic aspects}
\author{P P Goldstein$^{1}$, A M Grundland,$^{2,3}$ and S Post$^3$  }
\address{$^1$ Theoretical Physics Division, National Centre for Nuclear Research, Hoza 69, 00-681 Warsaw, Poland }
\address{$^2$ Department of Mathematics and Computer Sciences, Universit\'e du Quebec, Trois-Rivi\`eres. CP500 (QC) G9A 5H7, Canada }
\address{$^3$ Centre de Recherches Math\'ematiques. Universit\'e de Montr\'eal. Montr\'eal CP6128 (QC) H3C 3J7, Canada}

\ead{Piotr.Goldstein@fuw.edu.pl, grundlan@crm.umontreal.ca,
post@crm.umontreal.ca}
\begin{abstract}
In this paper, we consider both differential and algebraic
properties of surfaces associated with sigma models. It is shown
that surfaces defined by the generalized Weierstrass formula for
immersion for solutions of the $\cpn$ sigma model with finite
action, defined in the Riemann sphere, are themselves solutions of
the Euler-Lagrange equations for sigma models.  On the other hand, we show that the Euler-Lagrange equations for surfaces immersed in the Lie algebra $\su{n},$ with conformal coordinates, that are extremals of the area functional subject to a fixed polynomial identity are exactly the Euler-Lagrange equations for sigma models. In addition to these
differential constraints, the algebraic constraints, in the form
of eigenvalues of the immersion functions, are treated
systematically. The spectrum of the immersion functions, for
different dimensions of the model,  as well as its  symmetry
properties and its transformation under  the action of the ladder
operators are discussed. Another approach to the dynamics is
given, i.e. description in terms of the unitary matrix which
diagonalizes both the immersion functions and the projectors
constituting the model.

\end{abstract}

\section{Introduction}

The last few decades have seen important developments in the
investigation and construction of soliton surfaces associated with
integrable models. Their continuous deformations under various
types of dynamics have been the subject of extensive research, extending across the many areas of nonlinear physical
phenomena. In particular, the study of general properties of
nonlinear $\cpn$ sigma models and methods for finding associated
surfaces immersed in Lie algebras remains among the essential
subjects of investigation in several branches of mathematics and
physics.

The $\cpn$ models originated from the works of Gell-Mann and Levy
\cite{GellLevy1960}, in the middle of the previous century, in
order to explain the problem of the lifetime of charged pions by
introducing a new boson field. Later, this subject was treated by
Callan, Coleman, Wess and Zumino \cite{ callan1969structure,
coleman1969structure} in the low energy limit, where the authors
included  nonlinearity terms in the pion field $\phi.$ The main
feature of this approach is that the transformed pion field admits
a very simple Lagrangian density
\be \mathcal{L}=c\partial_\mu \phi^T \partial^\mu \phi, \label{Lphi} \ee
under the algebraic constraint
\be \phi^T\phi=\mathbb{I},\ee
where $\mathbb{I}$ is the $N\!\times\! N$ unit matrix and $c$ is some constant. The Lagrangian approach (\ref{Lphi})
proved to be a useful tool even in the case of a two-dimensional
domain since it appears in many areas of application in physics
(e.g. string theory \cite{PolStrom1991}, two-dimensional gravity
\cite{ GPWbook,Polybook}, quantum field theory \cite{NPWbook},
statistical physics \cite{Ou-YLuiXie1999}, fluid dynamics
\cite{DGZbook} etc), chemistry and biophysics (e.g. the
Canham-Helfrich membrane model \cite{Davbook, Landolfi2003}). This
subject has been generalized by many authors (e.g. \cite{Bobbook,
CarKon1996, Kono1996, KonoTaim1996, Sasaki1983, Uhlen1989,
Wardbook, ZakMik1979}) and more recently surveys of these
developments have been treated in several books (e.g.
\cite{Guestbook, Heleinbook2002, ManSutbook, Mik1986, Zakbook} and
references therein).

The differential algebraic approach to completely integrable
$\cpn$ sigma models in two dimensions and their associated
surfaces provides a rich class of geometric objects of study (see
e.g. \cite{GoldGrund2010, GoldGrund2011, GrundPost2011,
GrundPostZCR2012, GSZ2005, GrundPost2012, Kono1996, KonLand1999}).
In the description of the $\cpn$ model and all Grassmannian
models, it is more convenient to use projection operators as
variables, more specifically Hermitian projectors mapping onto
individual directions in $\cpn$ (rank-one) or on the appropriate
subspaces in Grassmannians (higher-rank). A Hermitian projection matrix
which maps onto a one-dimensional subspace $P(\xi, \overline{\xi})\in Aut(\C^N)$
satisfies
\be P^\dagger=P=P^2, \qquad \qquad \tr(P)=1.\nn \ee
The dynamics of the $\cpn$ sigma model defined on the Riemann
sphere are determined by the stationary points of the action
functional \cite{Zakbook}
\be S=\int _{\C} \tr\left[(\p P \bp P) +\mu (P^2-P) \right] d\xi d\overline{\xi},\quad \mu\in Aut(\C ^N),\label{S}
\ee
where the Lagrangian density is
\be\label{lagr-P} \mathcal{L} =\tr(\p P \bp P).\ee
The variation of the action $S$ yields the Euler-Lagrange (E-L) equations
\be\label{ELP} [\p\bp P, P]=0.\ee

It is a well known fact that any $\cpn$ solution of the E-L
equations (\ref{ELP}) with finite action, or equivalently
extendable to infinity, can be written in terms of raising and
lowering operators acting on a holomorphic or anti-holomorphic
solution respectively. This fact was first proven by Din and
Zakrzewski \cite{DinZak1980Gen} and later extended to ladder
operators for the projectors \cite{GoldGrund2010}
\be\label{raiselower} \Pi_+(P)=\frac{\p P P \bp P}{tr( \p P P \bp P)},\qquad \Pi_-(P)=\frac{\bp P P \p P}{tr( \bp P P
\p P)}.\ee The operators $\Pi_\pm$ map between finite action
solutions of the E-L equations and further, on this class of
solutions, the raising and lowering operators  are mutual inverses
and contracting operators \cite{GoldGrund2010, GrundPost2012}.
These facts imply that any rank-one Hermitian projector $P$ which
is a solution of the E-L equations can be written as a repeated
application of the raising operator on a holomorphic projector or,
equivalently, the lowering operator on an anti-holomorphic
projector. Recall \cite{GrundPost2012}, a holomorphic projector is
a projector which maps onto a direction in $\mathbb{C}P^{N-1}$
which has a holomorphic representative. This condition is
equivalent to the property that the projector be annihilated by
the lowering operator. The anti-holomorphic condition is
equivalent to the projector being annihilated by the raising
operator. Thus, any finite action solution of the E-L equations is a member
of some finite set of projector solutions of the E-L equations
defined by
\be P_k\equiv \Pi_+^kP_0 ,\qquad k=0,\dots, N-1\nn\ee
where $P_0$ is a holomorphic projector, i.e.
\be \Pi_-P_0=0.\ee
Note that these projectors are mutually orthogonal and, without loss of generality, provide a basis for $\cpn$
\be P_kP_j=\delta_{kj}P_j, \qquad P_k^\dagger=P_k, \qquad \sum_{j=0}^{N-1}P_j=\mathbb{I}.\ee

It is possible to express the E-L equations (\ref{ELP}) as a
conservation law \cite{GrunYurd2009}
\be \p [\bp P, P]+\bp[\p P, P]=0.\ee
This conservation law allows for the construction of a closed one-form
\be dX=i\left(-[\p P, P]d\xi +[\bp P, P]d\overline{\xi}\right).\ee
Hence, the integral
\be X(\xi, \overline{\xi})=i\int_{\gamma}\left(-[\p P, P]d\xi' +[\bp P, P]d\overline{\xi}'\right) , \qquad X^\dagger=-X,\label{Xint}\ee
 depends only on the endpoints $(\xi, \overline{\xi})$ of the curve $\gamma$ in $\C$ (i.e. it is independent of the trajectory in the complex plane), its other endpoint $(\xi_0, \overline{\xi}_0)$ is assumed to be fixed.
The function $X$ can be identified as a two-dimensional surface
immersed in a real $(N^2-1)$-dimensional Euclidean space. The
mapping $X:\mathbb{C} \ni (\xi, \overline{\xi})\rightarrow X(\xi,
\overline{\xi}) \in \su{N}$ is known in the literature
\cite{Kono1996} as the generalized Weierstrass formula for
immersion (GWFI) of two-dimensional surfaces in $\R^{N^2-1}\equiv
\su{N}.$

Consider now surfaces $X_k$ associated with finite action
solutions of the $\cpn$ sigma model given by the GWFI. The surface
is defined, up to affine transformations, by its tangent vectors
\be\label{Xdef} \p X_k=-i[\p P_k, P_k], \qquad \bp X_k=i[\bp P_k, P_k],   \ee
whose compatibility conditions are equivalent to the E-L equations (\ref{EL}).
The integrated form of the surfaces can be given explicitly \cite{GrunYurd2009}
\be \label{Xk}  X_k= -i\left(P _k+2\sum_{j=0}^{k-1}P _j-c_k \mathbb{I}\right), \qquad
c_k=\frac{1+2k}{N}.\ee As was shown in \cite{GrundPost2012}, the
surface $X_k$ (\ref{Xk}) is conformally parameterized and the
first fundamental form is proportional to the Lagrangian density.
Hence, the area of the surface is proportional to the action of
the physical model.

Finally, we mention that equation (\ref{Xk}) can be inverted to
solve for the projectors $P_k$ either as a linear combination of
the surfaces $X_1, \ldots, X_k$
\be
P_k=i\left(\sum_{j=1}^k(-1)^{k-j}(X_j-X_{j-1})+(-1)^kX_0\right)+\frac{1}{N}\mathbb{I}\label{PkXk},\ee
or by a nonlinear formula which depends on $X_k$ only
\cite{GoldGrund2010}
\be P_k=X_k^2-2i\left(\frac{2k+1}{N}-1\right)X_k-\frac{2k+1}{N}\left(\frac{2k+1}{N}-2\right)\mathbb{I}.\ee
The projective property $P_k^2=P_k$ then imposes a polynomial
constraint on the surfaces $X_k$. For any mixed solution of the
$\cpn$ model (\ref{ELP}) the minimal polynomial for the
matrix-valued function $X_k$ is the following cubic equation
\be \label{min} \left[X_k-ic_k\I\right]\left[X_k-i(c_k-1)\I\right]\left[X_k-i(c_k-2)\I\right]=\emptyset, \qquad 0<k<N-1.\ee
For holomorphic ($k\!=\!0$) or anti-holomorphic ($k\!=\!N\!-\!1$)
solutions, the minimal polynomials are quadratic
\be \label{minholo} \left[X_0-ic_0\I\right]\left[X_0-i(c_0-1)\I\right]=\emptyset,\ee
and
\be \label{minantiholo} \left[X_{N-1}-i(c_{N-1}-1)\I\right]\left[X_{N-1}-i(c_{N-1}-2)\I\right]=\emptyset,\ee
respectively \cite{GoldGrund2011-2, GrundPost2011}.

The main goal of this paper is to provide a self-contained,
comprehensive approach to such surfaces, namely two-dimensional
soliton surfaces associated with $\cpn$ sigma models immersed in
the $\su{N}$ Lie algebra. We begin our discussion by considering a
variational problem for the surfaces. We show that the variational
problem for the $\cpn$ sigma model is equivalent to a variational
problem for surface in conformal coordinates defined by the area
functional subject to a fixed polynomial identity. In particular,
surfaces defined by the GWFI for finite action solutions of the
$\cpn$ model defined on the Riemann sphere are conditional
extremals of this variational problem. We further show that
arbitrary immersion functions for a surface in $\su{N}$ in
conformal coordinates  which are extremals of the area functional
subject to some polynomials constraint $f_n(X)$ satisfy the same
E-L equations as the $\cpn$ model; namely,
\[ [\p \bp X, X]=0.\]
 Next we give a combinatorial
description of the distribution of these quantized eigenvalues and
the action of the raising and lowering operators on the spectrum.

The paper is organized as follows. In Section \ref{2}, we show
that the GWFI for surfaces associated with the $\cpn$ sigma model
with finite action, defined on the Riemann sphere, satisfy the E-L
equations. On the other other hand, we show that conformally
parameterized surfaces in $\su{N}$ that are extremals of the area
functional, subject to any polynomial identity, satisfy the same
E-L equations. Finally we propose another formalism for
description of the dynamics, namely a description in terms of a
unitary matrix $V$ which at the same time diagonalizes all the
surface immersion functions $X_k$ and the projectors $P_k$.
Section \ref{seceigen}, is devoted to the analysis of eigenvalues
of the GWFI for surface immersed in $\su{N}$. We establish
explicit formulae for the distributions of eigenvalues for
different dimensions, $N$ of the $\cpn$ model. The last section
contains remakes and suggestions regarding future developments.

\section{Euler-Lagrange equations for surfaces immersed in $\su{N}$}\label{2}
As described in Section 1, the surfaces defined by the GWFI are
conformally parameterized with the first fundamental form
proportional to the Lagrangian density and the surface area
proportional to the action of the model. In Subsection 1, we show
that the immersion functions themselves, $X_k,$  are conditional extremals of the area. We then consider general immersion functions into
$\su{N}$ that are subject to an arbitrary polynomial identity and
are stationary points of the area and show that such immersion
functions also satisfy the same E-L equations for the $\cpn$ sigma
model and in particular the surfaces $X_k$ satisfy this equation.
Another E-L formalism, in terms of the unitary matrix
diagonalizing both the projectors $P_k$ and the immersion
functions $X_k$, is given in Subsection 2.

\subsection{Direct Lagrangian description of the surfaces}

Recall that the the immersion functions $X_k$ defined as in
(\ref{Xk}) are conformally parameterized by $(\xi,
\overline{\xi})$ and their area is given by the action functional. Indeed, using the Killing form on $\su{N}$
\be \langle X, Y\rangle=-\frac12 \tr(XY), \qquad X, Y \in \su{N}\ee
as a metric on the tangent vectors to the surface, the area of the
surfaces, which are conformally  parameterized in the domain
$(\xi, \overline{\xi}) \in \Omega\subset \C$ is
\be A(X)=-\frac12\int_{\Omega} \tr (\p X \bp X) d\xi d\overline{\xi}.\ee

In the following proposition, we show that the the variational problem associated with the $\cpn$ sigma model is equivalent to a variational problem for surfaces in terms of their area. Furthermore,  surfaces defined by the GWFI for solutions of the $\cpn$ with finite action, defined on the Riemann sphere are conditional stationary points of such a variational problem.

\begin{proposition}\label{prop1}
Let $P$ be a rank-one Hermitian projector which is a solution of
the $\cpn$ sigma model with finite action, defined on the Riemann
sphere. Then, the immersion function $X$ defined by (\ref{Xint})
is a conditional stationary point of the area (which up to a
constant factor is also the action integral)
\be\label{action} A(X)=-\frac12\int_{\Omega} \tr (\p X \bp X) d\xi d\overline{\xi},\ee
under the condition $f(X)=0$, where $f$ is a given polynomial of at most
\iiird degree, with constant coefficients.
\end{proposition}
{\bf Proof}:
Recall \cite{GoldGrund2009}, that the immersion function $X$ defined as by (\ref{Xint}) with $P$ be a rank-one Hermitian projector which is a solution of
the $\cpn$ sigma model with finite action, defined on the Riemann
sphere, are conformally parameterized and so \eref{action} is the area of the surface.
From the definition of $X$ (\ref{Xint}) we have
\bea\label{dXdX}
\p X\bp X&=&[\p P,P][\bp P,P]\nn\\
&=&\p P P\bp P-P\p P\bp P P-\p P P^2\bp
P+P\p P P\bp P.
\eea
Using the properties of projectors
\be
\p P P=(\I-P)\p P,\quad \bp P P=(\I-P)\bp P,
\ee
which follow directly from differentiation of the projective
property $P^2=P$ \cite{GoldGrund2009,GoldGrund2010}, we see
 that the first and the last components of
(\ref{dXdX}) vanish, while the remaining two reduce to
\be\label{comcom}
-P\p P\bp P-(\I-P)\p P\bp P=-\p P\bp P.
\ee
But solutions of the $\cpn$ sigma models are conditional
stationary points of the action integral (\ref{S}), which consists
of integrated (\ref{comcom}) and the component corresponding to
the condition $P^2-P=0$.

It remains to show that this algebraic condition $P^2=P$ is equivalent to a fixed polynomial of at most  at most
\iiird degree, whenever $P$ is a finite action solutions of a $\cpn$ model, defined on the Riemann sphere. Under these conditions, the immersion function $X_k$ can be integrated (\ref{Xk}) and the projectors
$P_k,$ $k=0,...,N\! -\! 1,$ can be expressed as at most quadratic functions of the
corresponding $X_k$ (\ref{PkXk}). Then
the condition of (\ref{action}), $P_k^2-P_k=0$ is apparently \ivth
degree in $X_k$
\be
[X_k-i(c_k-1)\I]^2(X_k-i c_k\I)[X_k-i (c_k-2)\I]=0,
\ee
but the first factor is a total square, hence the minimal
polynomial of $X_k$ is the \iiird degree polynomial of
(\ref{min}).\qed

For the surfaces corresponding to holomorphic and antiholomorphic
solutions of the $\cpn$ models, it further reduces to a \iind
degree polynomial (\ref{minholo}) or (\ref{minantiholo})
respectively.

Conversely, let us consider an arbitrary smooth immersion function
for a surface $ X:\Omega\subset \C\rightarrow \su{N},$ where the
Lie algebra $\su{N}$ is realized as the set of anti-Hermitian
traceless $N\!\times\! N$ matrices. Suppose further that $X$ is a
conformal parameterization in terms of the complex variables $\xi$
and $\overline{\xi}, $
\be \tr(\p X \p X)=\tr(\bp X \bp X)=0. \nn\ee
The variational problem to be considered is concerning conditional stationary points of this functional with a constraint given by some polynomial equation
\be \label{characteristic equation} f_n(X)=\sum_{j=0}^n a_j X^j=0,\ee
with coefficients as smooth functions on $\Omega \subset \C$. This
choice of polynomial is equivalent to specifying the eigenvalues
of the matrix $X$. The extended functional, including a Lagrange
multiplier $\mu \in Aut(\C^N)$ is thus
\be A(X)=-\frac 12 \int_{\Omega} \tr\left( \p X \bp X
+\mu \left(\sum_{j=0}^n a_j X^j\right)\right) d\xi d
\overline{\xi}.\label{AL}\ee
A variation of the functional (\ref{AL}) yields the following E-L equations,
\be\label{E-Lsum} 2 \p \bp X+ \sum_{j=0}^n\sum_{k=0}^{j-1} a_j X^{j-1-k} \mu X^k=0.\ee
Taking (\ref{E-Lsum}) multiplied by $X$ on the left minus itself multiplied on the right gives
\be 2[X, \p \bp X] +\left(\sum_{j=0}^n a_j X^j\right)\mu-\mu \left(\sum_{j=0}^n a_j X^j\right),\ee
which reduces to the Euler-Lagrange equation
\be [\p\bp X, X]=0, \label{EL} \ee
modulo the characteristic equation (\ref{characteristic equation}). We thus have the following proposition.

\begin{proposition} \label{prop2}  Let $f_n(\cdot)$ be a polynomial
with coefficients that vary smoothly on some domain $\Omega
\subset \C$. Consider the set of all smooth immersion functions
from $\Omega$ to the Lie algebra $\su{N}$ which are parameterized
by conformal coordinates $\xi$ and $\overline{\xi}$ and which
fulfill the requirement $f_n(X)=0$. The elements of this set that
are extremals of the area functional satisfy the Euler-Lagrange
equation (\ref{EL}).
\end{proposition}

It is a direct consequence of the projective property of the set
of orthogonal projectors that the surface $X_k$ also satisfies the
E-L equations (\ref{EL}).

\begin{corollary}
Let $P$ be a rank-one Hermitian projector which is a solution of
the $\cpn$ sigma model with finite action, defined on the Riemann
sphere. Then, the immersion function $X$ defined by (\ref{Xint}) satisfies (\ref{EL})
\end{corollary}
{\bf Proof:} As mentioned above, the immersion functions $X$ defined by (\ref{Xint})  for $P$ be a rank-one Hermitian projector which is a solution of
the $\cpn$ sigma model with finite action, defined on the Riemann
sphere are conformally parameterized and satisfy an at most  \iiird degree polynomial. Thus, they are conditional stationary points  of the variational problem defined by the area functional and its minimal polynomial. The previous proposition therefore says that the functions $X$ satisfy the E-L equations \eref{EL}. \qed

\subsection{Description in terms of the unitary diagonalizing matrix}

Since the projectors are Hermitian matrices, they are
diagonalizable by a unitary transformation. Being orthogonal to
one another, they commute and thus a common diagonalizing unitary
matrix exists for all of them. The same matrix diagonalizes also
the surface immersion functions $X_k$ as they may be expressed as linear
combinations of the projectors $P_k$ and the unit matrix (\ref{Xk}).
Moreover, the eigenvalues of the projectors $P_k$ are constants: 0 or 1,
hence it is the diagonalizing  matrix, which contains the whole
dynamics including differential properties of all the projectors
and surfaces. Therefore it seems worthwhile to find the equations
which govern its dynamics and to derive the corresponding
Lagrangian formalism.

Let $V(\xi^1,\xi^2)$ (where $\xi=\xi^1+i\xi^2$) be a unitary matrix such that
\be
V^\dagger P_k V = \I_k,~\mathrm{~for~all}~ k=0,...,N\!-\! 1,\label{V}
\ee
where $\I_k$ is a diagonal matrix consisting of $1$ in the $k$-th
place and zeros otherwise. Then the dynamics of $V$ can be derived
from the equations of the dynamics at any other level e.g. from the
dynamics of projectors (\ref{ELP}) or directly from the equation
of dynamics in terms of the inhomogeneous normalized variables
$z(\xi^1,\xi^2)$, with $P=z\otimes z^\dagger$,  which reads
\be\label{z-eqs1}
(\I - z\otimes\zd)[\p_\mu\p_\mu z-2(\zd\p_\mu z)\p_\mu z]=0, \qquad z^\dagger z=1,
\ee
(the summation convention has been assumed for repeating Greek indices). Equation (\ref{z-eqs1}) has a compact form in terms of the
covariant derivative \cite{Zakbook}
\be\label{cov}
D_\mu\psi = \p_\mu\psi-(\zd\p_\mu z)\psi,
\ee
namely
\be\label{z-eqs}
D_\mu D_\mu z+(D_\mu\zd D_\mu z)z=0,\qquad z^\dagger z=1.
\ee
From the equivalent equations (\ref{z-eqs1}), (\ref{z-eqs}) and
(\ref{ELP}), we choose the last one as a starting point for our
derivation. The equation (\ref{ELP}) in terms of the diagonalizing
matrix reads
\be
[\p_\mu\p_\mu (V\I_k\Vd),\, V\I_k\Vd]=0.
\ee
We perform the differentiation replacing all the derivatives of
$\Vd$ by the corresponding derivatives of $V$, using the identity
\be\label{diff-prop}
\Vd V=V\Vd = \I~,\mathrm{~whence~}\p_\mu\Vd=-\Vd\p_\mu V\Vd.
\ee
Making use of the property of $\I_k$
\be
\I_k A\I_k=A_{kk} \I, \quad ~\mathrm{for~any~N\!\times\! N~matrix}~A,
\ee
we obtain (after a right-multiplication by $V$) an intermediate
product, from which two different nontrivial \iind order equations
can be derived. The intermediate product reads
\bea\label{intermediate}
&&-\p_\mu\p_\mu V\I_k-V\I_k\Vd\p_\mu\p_\mu
V+2(\Vd\p_\mu\p_\mu V)_{kk}V\I_k\nn\\
&&+2V\I_k\Vd\p_\mu V\Vd\p_\mu V - 2(\Vd\p_\mu V\Vd\p_\mu
V)_{kk}V\I_k\nn\\ &&-2(\Vd\p_\mu V)_{kk}V\I_k\Vd\p_\mu
V+2(\Vd\p_\mu V)_{kk}\p_\mu V\I_k=0.
\eea
The first form of the $V$-dynamics equation is obtained by summing
up equations (\ref{intermediate}) for $k=0,...,N\! -\! 1$ over
$k$. Bearing in mind that $\sum\,\I_k=\I$ (the unit matrix), we
get (after dividing by $-2$)
\bea\label{1st-version}
&&\p_\mu\p_\mu V- V\diag(\Vd\p_\mu\p_\mu V)- \p_\mu
V\Vd\p_\mu V+V\diag\left((\Vd\p_\mu V)^2\right)\nn\\
&&-\p_\mu V\diag(\Vd\p_\mu V)+V\diag(\Vd\p_\mu V)\Vd\p_\mu V=0,
\eea
(for a square matrix $B$, $\diag(B)$ denotes its diagonal part,
with zeros outside the diagonal).

Equation (\ref{1st-version}) may be cast into a compact form if we
extend the definition of the covariant derivative to square
matrices, namely:
\be\label{cov-matr}
D_\mu A = \p_\mu A-A\,\diag(\Vd\p_\mu V).
\ee
 This definition is consistent with the
definition of the covariant derivative for vectors (\ref{cov}),
namely if we regard the matrix $A$ as built of column vectors,
then the columns of its covariant derivative (\ref{cov-matr}) are
covariant derivatives of its columns, in the usual sense
(\ref{cov}). In terms of the covariant derivatives
(\ref{cov-matr}), equation (\ref{1st-version}) reads simply as
\be\label{cov-form1}
D_\mu D_\mu V-(D_\mu V)\Vd D_\mu V=0.
\ee
Another nontrivial form of the dynamic equation for $V$ may be
obtained if we right-multiply each of the equations
(\ref{intermediate}) for $k=0,...,N\!-\!1$ by $\I_k$ before
summing them over $k$. Then the result reads
\bea\fl \label{2nd-version}
\p_\mu\p_\mu V-V\diag(\Vd\p_\mu\p_\mu V)-2\p_\mu
V\diag(\Vd\p_\mu V)+2V\,[\diag(\Vd\p_\mu V)]^2=0.
\eea
Equation (\ref{2nd-version}) is naturally obtained if we start
from the $z$-equation (\ref{z-eqs}). To get a compact form of this
equation we define for a square matrix $A$
\be\label{var}
\var(A)=\diag(A^2)-[\diag(A)]^2,
\ee
which allows us to write equation (\ref{2nd-version}) as
\be\label{covar-form}
D_\mu D_\mu V-V\var(\Vd\p_\mu V)=0.
\ee
i

It will be clear from the Lagrangian approach to the dynamics that
the version (\ref{1st-version}) contains both the information on
the orthogonality of the projectors $P_k$ and their projective
property, while equation (\ref{2nd-version}) only that of the
normalization of the vectors $z_k$, which is equivalent to the
projective property of $P_k$ but it does not impose their
orthogonality.

Equations (\ref{1st-version}) may easily be derived from the
Lagrangian density
\bea\label{lagr-v1}
\mathcal{L}&=& \tr\left[\left(D_\mu V\right)^\dagger D_\mu
V-\nu\Vd V\right]\nn\\&=&\tr\left[\p_\mu \Vd\p_\mu V
-\diag\left(\p_\mu\Vd V\right)\diag\left(\Vd\p_\mu
V\right)\right],
\eea
where $\nu\in Aut(\C ^N)$ is the Lagrange multiplier corresponding
to the unitarity condition $\Vd V=\I$. The multiplier $\nu$ may be
restricted to Hermitian matrices as both $\left(D_\mu
V\right)^\dagger D_\mu V$ and $\Vd V$ are Hermitian (note that for
Hermitian $\nu$, the \iind component $\tr(\nu\Vd
V)=\tr\left[(1/2)(\nu\Vd V+\Vd V\nu)\right]$ is a trace of a
Hermitian operator).

The E-L equations corresponding to the Lagrangian density
(\ref{lagr-v1}) read
\bea\label{lagr-nu1}
&&-\p_\mu\p_\mu V+ V\diag(\Vd\p_\mu\p_\mu V)-2V\diag\left( (\Vd\p_\mu V)^2\right)\nn\\
&&+2\p_\mu V\diag(\Vd\p_\mu V)-\frac12 V\nu=0.
\eea
Left multiplication by $2\Vd$ allows us to extract the Lagrange
multiplier $\nu$. Under the assumption that $\nu$ is Hermitian,
similar operations on the Hermitian conjugate of (\ref{lagr-nu1})
yield the same matrix $\nu$. The condition that both expressions for
$\nu$ are equal, it is exactly equation (\ref{1st-version}),
provided that we express the $\p_\mu\p_\mu\Vd$ and $\p_\mu\Vd$ in
terms of the derivatives of $V$ (by means of (\ref{diff-prop}) and
its derivative with respect to $\xi_\mu$).

The diagonalizing matrix may be constructed out of the
inhomogeneous variables $z_k$. Namely, the matrix $V$ takes the form
\be\label{matrixV}
V= \left(\ba{lll}z_0^0&...&z_{N-1}^0\\
                \dotfill&...&\dotfill\\
                z_0^{N-1}&...&z_{N-1}^{N-1}
\ea\right),
\ee
where the lower index at $z_0,z_1...$ numbers the vectors $z_k$
while the upper one numbers the components of each vector.

 Substituting the above form of $V$ in the Lagrangian
(\ref{lagr-v1}), we obtain
\be\label{lagr-z}
\mathcal{L}=\sum\limits_i \p_\mu z_i(\I-z_i\otimes\zd_i)\p_\mu
z_i-\sum\limits_{i,j}\nu_{ij}\zd_j z_i,
\ee
i.e. the Lagrangian density is a sum of the Lagrangian densities
\be\label{lagr-zi}
\mathcal{L}_i=\p_{\mu} \zd_i(\I-z_i\otimes\zd_i)\p_\mu z_i-\sigma
\zd z,
\ee
except for the constraint term $\sigma \zd z$, which imposes weaker
constraints than those in (\ref{lagr-z}). The constraints in
(\ref{lagr-z}) contain both normalization and orthogonality of
$z_i$, while those of (\ref{lagr-zi}) -- only the normalization.
If we sum up the Lagrangians (\ref{lagr-zi}) for $i=0...N\!-\!1$,
then the constraint term may be written as
$\diag(\Vd)\diag(V)=\I$, which requires only a diagonal matrix as
the Lagrange multiplier, $\sigma$. In such a case, we do not have to
subtract the Hermitian conjugate of  (\ref{lagr-nu1}). The
multiplier $\nu$ may be obtained by simple taking the diagonal
part of (\ref{lagr-nu1}) to give
\be
\nu=4\,\left\{ [\diag(\Vd\p_\mu V)]^2-\diag\left((\Vd\p_\mu
V)^2\right)\right\}=-4\,\var(\Vd\p_\mu V).
\ee
Substitution of this $\nu$ to  (\ref{lagr-nu1}) yields exactly
the \iind form of the equation governing the dynamics of $V$,
namely (\ref{2nd-version}).

Summarizing, equation (\ref{1st-version}), which governs the
dynamics of the diagonalizing matrix  $V$, has built-in
orthogonality of the $\cpn$ vectors $z_k$, because they are
columns of a unitary matrix and the unitarity condition is imposed
in the Lagrangian as a constraint (\ref{lagr-v1}). In this aspect
the description in terms of the diagonalizing matrix differs from
all the previous descriptions (in terms of the inhomogeneous
vectors $z_k$, the homogeneous vectors $f_k$ or the projectors
$P_k$) where the orthogonality was proven.

\section{Eigenvalues of the generalized Weierstrass immersion function for
surfaces}\label{seceigen}
It can be seen from the above discussion, that the choice of the
eigenvalues for the immersion functions (or equivalently the
minimal polynomials) is fundamental to the model described. In the
following section, we shall investigate the eigenvalues in depth
including their distribution and the action of the ladder
operators on the spectrum.

\subsection{Eigenvalues of the immersion functions $X_k$ for a fixed $N$}

In this section, we fix $N$, the dimensions of the space $\cpn,$
and consider the eigenvectors of the immersion functions $X_k$.
The decomposition of unity
\be \mathbb{I}=\sum_{j=0}^{N-1} P_j\ee
is used to express the surfaces as  linear combinations of the eigenvectors \cite{GoldGrund2011}
\be\label{XkP} X_k= -i\left((2-c_k)\sum_{j=0}^{k-1}P _j+(1-c_k)P _k-c_k\sum_{l=k+1}^{N-1} P_l\right)\in \su{N}.\ee
The corresponding eigenvalues of the immersion functions $X_k$ take one of the following forms
\bea
i.) & \quad ic_k=\frac 1 N +\frac{2k}{N},& \quad k=0,1,.., N-2,\nn \\
ii.) & \quad i(c_k-1)=\frac{1-N}{N}+\frac{2k}N, & \quad k=0,1,..., N-1,\nn\\
iii.) & \quad i(c_k-2)=\frac{1-2N}{N}+\frac{2k}N, & \quad k=1,..., N-1.\nn\eea
Thus, for each $k,$ the eigenvectors are given by the set $P_j$ with eigenvalues
\be X_k P_j=i\lambda P_j, \qquad \lambda= \left\{ \ba{cc} c_k-2 & j<k,\\ c_k-1 & j=k,\\ c_k & j>k.\ea\right.\ee
Note that this implies that $[X_k,P_j]=0$ as can be directly
observed from the form of $X_k$ given in (\ref{XkP}). The
eigenspaces, for a fixed $k,$ are
\bea E_{c_k-2}=\lbrace P_0e,\ldots, P_{k-1}e\rbrace,\nn \\
  E_{c_k-1}=\lbrace P_ke \rbrace,\nn \\
  E_{c_k}=\lbrace P_{k+1}e,\ldots, P_{N-1}e\rbrace, \eea
where $e$ is any unit vector not orthogonal to any $P_k.$ Note
that for $k=0,$ $E_{c_0-2}=\emptyset$ and for $k=N-1,$
$E_{c_{N-1}}=\emptyset.$ The degeneracy of $c_k$ is $N-k-2,$ that of
$c_k-1$ is 1 and that of $c_k-2$ is $k.$ Let us define
\be S_k=\lbrace c_k, c_k-1, c_k-2\rbrace \ee
to be the set of eigenvalues of $-iX_k.$ Here we have normalized
out the constant $i$ for each eigenvalue. Also, note that $P_je$
is an eigenvector for each $X_k,$ though generally with different
eigenvalues.

Recall, from \cite{GoldGrund2010}, that there exist raising and
lowering operators $\chi_\pm$, based on the raising and lowering
operators (\ref{raiselower}) which are maps from the surface $X_k$
to $X_{k\pm1}$
\be \chi_\pm (X_k) =X_k\mp \left(\Pi_\pm (P_k)+P_k\right)\pm \frac{2i}N\mathbb{I}.\ee
Repeated application of the raising and lowering operators
$\chi_\pm$ generates the union of the set of eigenvalues for each
of the surfaces $X_k$. That is, let $S(N)$ be the union of the set
of eigenvalues for each $X_k$ for $k=0, \ldots, N-1$ for a fixed
$N$. Here, we denote $S(N)=S$ where the dimension $N$ is fixed.

Starting from $X_0,$ the surface has two eigenvalues $c_0-1$ and
$c_0.$ In order to generate the next set of eigenvalues, we put a
``ghost'' eigenvalue at $c_0-2.$ To obtain the next set of
eigenvalues, we simply raise each eigenvalue by $2/N$ and the
ghost eigenvalue becomes real. The surface $X_1$ thus has three
distinct eigenvalues. The interpretation of the ``ghost''
eigenvalue is that it is the pre-image of $c_1-2$ under the
mapping $\chi_+$ or alternatively it is the annihilated image of
$c_1-2$ under the mapping $\chi_-.$

If we continue the procedure $N-1$ times, we obtain the set
\be S(N)=S=\cup_{k=0}^{N-1} S_k.\ee
In this process, the highest eigenvalue $c_{N-1}$ is also
annihilated and so produces another ``ghost'' eigenvalue at the
opposite end of the spectrum. Note that the procedure could have
begun at the top of the spectrum by applying the lowering operator
to the eigenvalues of $X_{N-1}$ plus a ghost dot at the top of the
spectrum.  Figure 1 shows this process where we denote $c_k$ and
$c_k-2$ by dots, $c_k-1$ by $x$'s and the action of $\chi_+$ by
arrows at the top. The ``ghost'' dots at either end of the
spectrum are filled in.

\begin{figure}\label{chi}
\begin{center}
\includegraphics[width=1.0\textwidth]{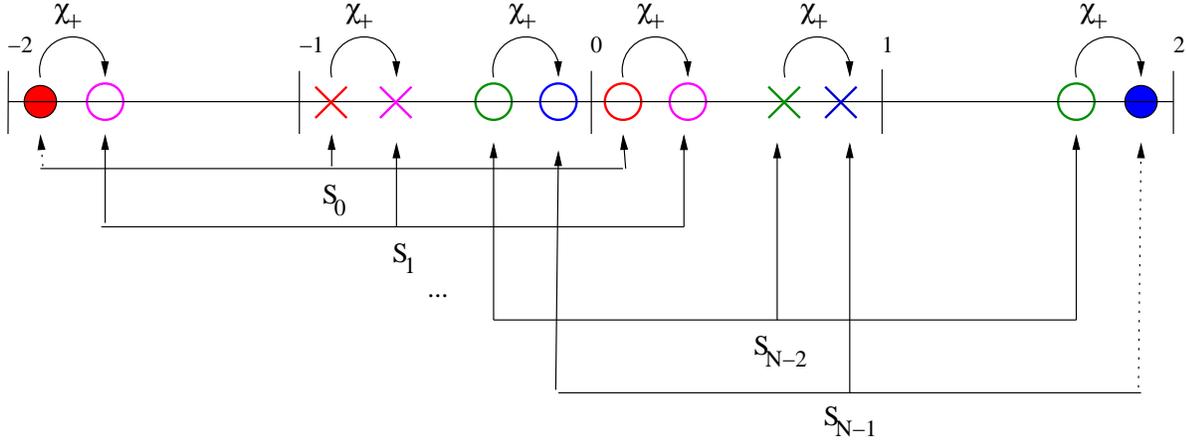}
\end{center}
\caption{{\bf Action of $\chi_\pm$ on the eigenvalues}: $\chi_\pm$
is the mapping between surfaces $X_{k\pm 1}.$ Its action is
denoted by the arrows. The filled circles are ``ghost''
eigenvalues and the empty circles correspond to eigenvalues which
are realized for some $k$. Here we have omitted the sets $S_2$
through $S_{N-2}$ as denoted by the ellipses.}
\end{figure}

Let us consider the set $S$ of all eigenvalues of immersion
functions, $X_k$ for $k=0,...,N-1$ for a fixed $N.$  Further, let
us define the following sets
\bea \ba{l} D^+\equiv \lbrace c_k\rbrace =\left\{\left. \frac 1 N +\frac{2k}{N}\right| k=0,1,.., N-2\right\},\\
  D^0\equiv \lbrace c_k-1\rbrace =\left\{\left. \frac {1-N} N +\frac{2k}{N}\right| k=0,.., N-1\right\},\\
D^-\equiv \lbrace c_k-2\rbrace =\left\{\left. \frac {1-2N} N +\frac{2k}{N}\right| k=1,.., N-1\right\},\ea \eea
and
\bea
  D\equiv D^+\cup D^-, \qquad S\equiv D\cup D^0.\eea

\begin{proposition}
For the sets $D^\pm$ and $D^0$ defined above, we have the following restrictions, for all $N$
\bea  D^+ \subset [\frac1N, 2-\frac3N],\nn\\
 D^0 \subset [-1+\frac 1N , 1-\frac1N],\nn\\
 D^-\subset [-2+\frac3N, -\frac 1N].\nn \eea
\end{proposition}
{\bf Proof:} These follow from direct observations. The lowest
element of $D^+$ is achieved for $k=0$ and has a value of  $\frac
1N$ whereas the highest element is achieved for $k=N-2$ and has
the value of $2-\frac3N.$ The lowest element of $D^0$ is achieved
for $k=0$ and has a value of  $-1+\frac 1N,$ whereas the highest
element is achieved for $k=N-1$ and has the value of $1-\frac1N.$
The lowest element of $D^-$ is achieved for $k=1$ and has a value
of  $ -2+\frac1N$ whereas the highest element is achieved for
$k=N-1$ and has the value of $-\frac1N.$ Also note that\[S=D^+\cup
D^0\cup D^-\subset [-2+\frac 3N, 2-\frac3N],\] and the distance
between the highest and lowest eigenvalue is $d=2(2-\frac3N).$
\qed

\subsubsection{Symmetries of the sets $D^\pm$ and $D^0$}

Here we shall prove that $D^0$ is symmetric about the origin and
that the image of $D^+$ under multiplication by $-1$ is $D^-,$
i.e. for any $x\in D^+,$ $-x\in D^-.$
\begin{proposition}
$D^0$ is symmetric about the origin.
\end{proposition}
{\bf Proof}: Let $x \in D^0$. Then
\bea x&=&\frac{1-N+2k}{N},\nn\\
 -x
  &=& \frac{1-N-2l}{N},\qquad\nonumber l=N-k-1.\eea
  This implies that $-x\in D^0.$ \qed

\begin{proposition}
For any $x$ in $D^+,$ $-x\in D^-.$
\end{proposition}
{\bf Proof}: Let $x \in D^+.$
\bea x&=& \frac{2k+1}N,\nn\\
-x 
 &=& \frac{1-2N+2l}{N}, \qquad l=N-k.\nn \eea
 This implies that $-x \in D^-$ \qed

\subsubsection{Intersections of the sets $D^\pm$ and $D^0$}
 Here we claim that, for $N$ even, $D^0\subset D$ and, for $N$ odd, $D\cap D^0=\emptyset.$
 \begin{proposition}\label{odddist}
 For $N$ odd, $D^0\cap D=\emptyset.$
 \end{proposition}
 {\bf Proof}:  Let $x \in D^0$ and $N=2n+1,$  then
   \[ x=\frac{1+2k-2n-1}{2n+1}.\]
Now, if $x\in D^+$ there exists some $k$ such that
\[ \frac{2k-2n}{2n+1}=\frac{1+2k}{2n+1},\]
which is impossible because of the parity of the numerators.
On the other hand, for $x$ to be in $D^-$ we require
\[\frac{2k-2n}{2n+1}=\frac{1+2k-4n-2}{2n+1},\]
which is also a contradiction because of the parity of the
numerators. Thus, $D^0\cap D=\emptyset.$ \qed

 \begin{proposition}\label{evendist}
 For $N$ even, $D^0\subset D$
 \end{proposition}
 {\bf Proof}: Suppose that $N$ is even, i.e. $N=2n$ with $n=1....$ Let $x\in D^0,$
 then \bea x&=&\frac{1-2n+2k}{2n},\qquad k\in \{0,..,2n\} .\nn \eea
Suppose also, without loss of generality, that $x\ge0$ then $k$
must be greater than $n$ and so it is possible to define $l=k-n$.
Since $n \ge k\leq 2n-1,$ $0\ge l=k-n\leq 2n-2$ and so \[ x=
\frac{1+2l}{2n}, \qquad l=k-n\in \{0,..,2n-2\} .\]
 That is, $x\in D^+.$
By the symmetry arguments in Section 3.2, if $x$ is negative then
$x\in D^-$ and so in either case $x\in D=D^+\cup D^-.$ \qed

\subsubsection{Spacing between elements of $S$}
It can be directly observed from the definition of the $c_k$'s
that the difference between each $c_k$ is $2/N.$ Thus the spacing
between elements in $D^+$, $D^-$ and $D^0$ is always $2/N.$
\begin{proposition}
For $N$ even, the step size between elements of $S$ is $2/N$.
\end{proposition}
{\bf Proof}: From Proposition \ref{evendist}, $D^0\subset D$ and
so $S=D.$ Further, we know that $D^+\subset[\frac1N, 2-\frac3N]$
and $D^-\subset[-2+\frac3N, -\frac1N]$ and so these sets are
disjoint and the elements within each of them are at a distance
$2/N$ apart. Finally, the highest element of $D^-$ is $-1/N$ and
the lowest element of $D_+$ is $1/N$ and so they are at a distance
of $2/N$ apart. Thus, all elements of $S$ are at a distance of
$2/N$ from their nearest neighbor. \qed

\begin{proposition}
For $N$ odd, on the interval $[-1,1]$, the step size between
elements of $S$ is $1/N.$ On the intervals $[1,2-\frac3N]$ and
$[-2+\frac3N, -\frac1N]$ the step size is $2/N.$
\end{proposition}
{\bf Proof}: First we show that, for $N$ odd, $1\in D^+$ and
$-1\in D^-.$ Assume that $N=2n+1, $ then
\bea 1=\frac 1 {2n+1} +\frac{2n}{2n+1}=c_n\in D^{+}, \nn\\
-1=1-2=c_n-2\in D^-.\nn\eea
Thus, for $N=2n+1$ odd,
\bea D\cap [-1,1]&= \left\{-1, -1 +\frac{2}{2n+1},..., 1-\frac{2}{2n+1},1\right\} \nn\\
&=\left\{\left.-1+\frac{2\ell}{2n+1}\right|\ell=1,...,2n+1\right \}\nn.\eea
On the other hand
\bea D^0=\left\{ \left.-1+\frac{2\ell+1}{2n+1}\right|\ell=1,...,2n+1\right\},\nn\eea
and so, for $N$ odd,
\bea S\cap [-1,1]&=&(D\cap [-1,1])\cup D^0\nn\\
&=&\left\{\left.-1+\frac{k}{2n+1}\right|k=1,...,2n+1\right\},\nn\eea
and in particular, the difference between the energy values is $1/N.$

However, on the intervals $[1,2-\frac3N]$ and $[-2+\frac3N,
-\frac1N],$ $S$ is comprised solely of $D^+$ and $D^-$
respectively and so the the step size is $2/N$ as remarked above.
\qed

\subsubsection{Counting elements of S}
 Define $|A|$ to be the number of elements in the set, then the following identities hold:
\bea D^+\cap D^-=\emptyset, \qquad |D^+|=N-1, \quad |D^-|=N-1.\nn\eea
We can immediately see that:
\begin{enumerate}
\item If $N$ is odd,
\bea |S|=|D|+|D^0|=2N-2+N=3N-2.\nn\eea
\item If $N$ is even,
\bea |S|=|D|=2N-2.\nn\eea
\end{enumerate}
Note that, counting repeated eigenvalues, the number of
eigenvalues is $3N-2$. For $N$ odd, as was proven in Proposition
\ref{odddist},  no eigenvalues are repeated and so there are
$3N-2$ distinct eigenvalues. For $N$ even, as was proven in
Proposition \ref{evendist}, all of the values in $D^0$ are also in
$D$ and so there are $N$ repeated eigenvalues (i.e. shared by two
different surfaces) and so only $2N-2$ are distinct.

\subsubsection{Graphs}
Let us display the eigenvalues of $X_k$ for $k=0,\ldots, N-1$,
i.e. the set  $S=D\cup D^0$, on the real line in the following
graphs, figures 2 and 3. Here we represent the elements of $D$ by
dots and the elements of $D^0$ by $x$'s. We also include the two
``ghost'' values of dots by filled in circles. Note that the set
of points distributed along the real line is symmetric with
respect to the origin. Therefore, it is sufficient to consider
only the right or left half of the set $S$.

\begin{figure}\label{even}
\begin{center}
\includegraphics[width=1.0\textwidth]{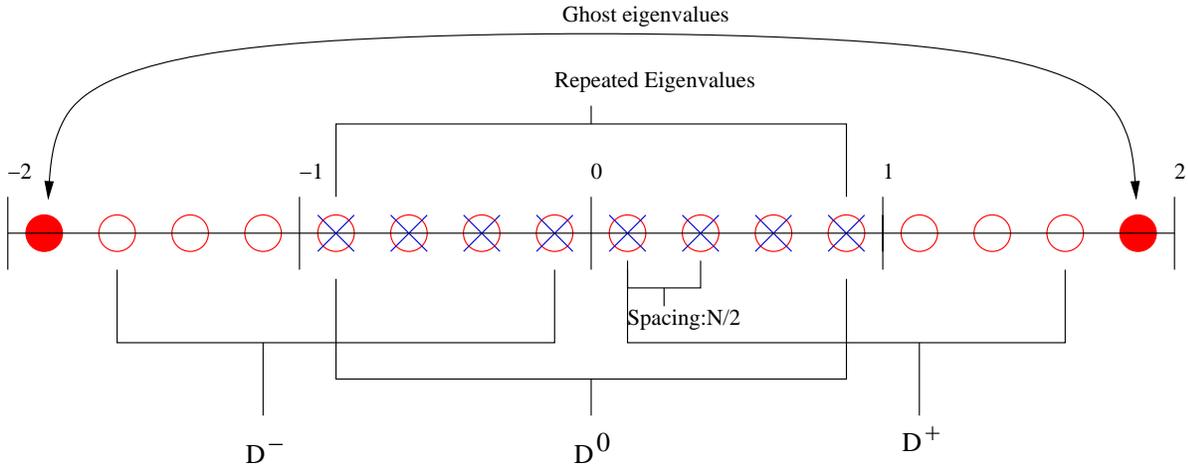}
\end{center}
\caption{{\bf Eigenvalues for $N=8$ (even)}: The elements of $D$
are represented by hollow dots and the eigenvalues of $D^0$ are
represented by $x$'s. The filled in circles represent ghost
eigenvalues.} \label{figeven}
\end{figure}

\begin{figure}\label{odd}
\begin{center}
\includegraphics[width=1.0\textwidth]{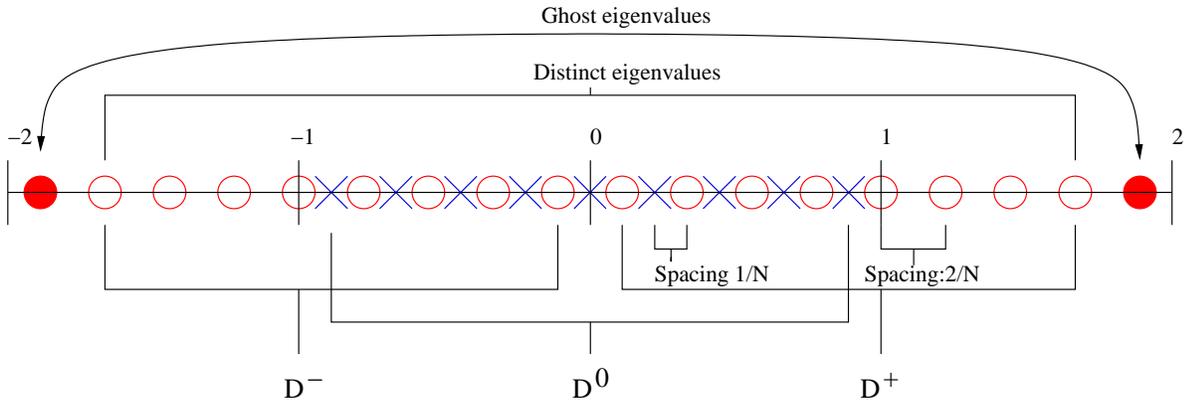}
\end{center}
\caption{{\bf Eigenvalues for $N=9$ (odd)}: The elements of $D$
are represented by hollow dots and the eigenvalues of $D^0$ are
represented by $x$'s. The filled in circles represent ghost
eigenvalues.} \label{figodd}
\end{figure}

\subsection{Mapping between sets of eigenvalues for different $N$}
The question to be answered in this section is whether there
exists an  operator $A_\pm$ such that
\be A_+S(N)=S(N+1), \qquad A_-S(N)=S(N-1), \nn \ee
which lowers or raises the number of dots? Here the answer is affirmative.

Recall, the eigenvalues of each surface $X_k$ depend on the
dimension of the space $N$
\be S_k(N)=\{c_k,c_k-1,c_k-2\}, \qquad c_k=\frac{1+2k}{N}, \nn\ee
and hence the union of these values depends on $N$ not only in the
summation but also in the values of the individual $S_k$'s i.e.
\be S(N)=S=\bigcup_{k=0}^{N-1} S_k(N).\nn\ee
In this section, we will outline a simple procedure for performing
the induction on $N$ which maps from $S(N)$ to $S(N\pm1)$. Let us
consider the sets $D^\pm$ and $D^0$ and describe them in a way
that is conducive to performing induction on $N$. We denote
$D^\pm=D^\pm(N)$ and $D^0=D^0(N)$ to make explicit the dependence
on $N$.

We describe each set in the following way. For each set, we place
in the interval $M$ points such that
\begin{itemize}
 \item[1).] the points are evenly spaced throughout the interval at a maximal
 distance
\item[2).]
the points are at least a distance $1/M$ from the end points of
the interval where $M$ is the number of points in the interval.
\end{itemize}
Thus, $D^-(N)$ is the set of $N$ points satisfying conditions 1
and 2, where $M=N,$ on the interval $[-2,0]$ where the lowest
point is designated a ``ghost'' point. Similarly, $D^+(N)$ is the
set of $N$ points satisfying conditions 1 and 2 on the interval
$[0,2]$ where the highest point is designated a ``ghost'' point.
The set $D^0(N)$ is the set of $N$ points on the interval $[-1,1]$
satisfying conditions 1 and 2. Then \[S(N)=D^-(N)\cup D^0(N)\cup
D^+(N).\]

The same conditions hold for $M=N+1.$ That is $D^-(N+1)$ is the
set of $N+1$ points satisfying conditions 1 and 2 on the interval
$[-2,0]$ where the lowest point is designated a ``ghost'' point.
Similarly, $D^+(N+1)$ is the set of $N+1$ points satisfying
conditions 1 and 2 on the interval $[0,2],$ where the highest point
is designated a ``ghost'' point. The set $D^0(N+1)$ is the set of
$N+1$ points on the interval $[-1,1]$ satisfying conditions 1 and
2.
 Then
 \[S(N+1)=D^-(N+1)\cup
 D^0(N+1)\cup
 D^+(N+1).\]

With this description we can immediately see the induction step.
To describe $S(N+1)$ in terms of the configurations of $S(N)$, we
need only add a point to each interval and require the same
conditions, 1 and 2,  on the arrangement of the points.  For the
set $S(N-1)$ we subtract one point from each interval and require
that 1 and 2 hold.

Note that this description does not depend on the parity of $N$
and in particular, the points in each set may overlap, depending
on the parity of $N$, and result in repeated eigenvalues.

 \subsection{Completeness of the set of eigenvalues}

 It is apparent that the surfaces $X_k,~k=0,\dots , N\! -\! 1$ are too few to make
 a basis in the Lie algebra $\su{N}$. However, for each $(\xi, \overline{\xi})\in \Omega \subset \C,$ we have the following result,
\begin{proposition}
\begin{enumerate}
 \item The vector subspace spanned by the surfaces $X_k~ (k=0,\dots N\! -\! 1)$,
 exhausts all the possible spectra for elements of the Lie algebra $\su{N}$.
\item  The projectors $P_k$ are not traceless, therefore they span
 a larger subspace, whose spectra may assume all the values from
 $\mathbb{C}^N$.
\end{enumerate}
\end{proposition}
{\bf Proof}: The surfaces $X_k$ as well as the projectors $P_k$ are
diagonalizable (the first ones as antihermitian matrices, the
second ones as hermitian matrices). As before, let $V$ be the
diagonalizing matrix for the projectors (\ref{V}) and let
$\lam_0,...,\lam_{N-1}$ be an arbitrary sequence of complex
numbers. Consider a linear combination of the projectors
\[S=\sum_{k=0}^{N-1}\lam_k P_k=V^{-1}\left(\sum_{k=0}^{N-1}\lam_k\mathbb{I}_k\right)V.\]
The matrix S has eigenvalues $\lambda_0,\ldots \lambda_N$ and so
this completes the proof of the part part {\it (ii)} of the
proposition.
 Part part {\it (i)} follows
from the fact that each of the projectors $P_k$ may be expressed
as a linear combination of the traceless matrices $X_k,~ k=1,\dots
, N\! -\! 1$ and the unit matrix $\mathbb{I}$ (\ref{PkXk}). Thus
the matrix $S$ can be expressed as a linear combination of the
$X_k$'s and the unit matrix.  The eigenvalues satisfy
$\lambda_0+\ldots\lambda_{N-1}=0$ if and only if $S$ is traceless
and so the coefficient of $\mathbb{I}$ is zero. Hence the matrix
$S$ is a linear combination of $X_k$ and thus it is an element of
the subalgebra spanned by the $X_k$. \qed
Similarly, we have the following result.
\newtheorem{conjugate}{Theorem}
\begin{proposition}\label{prop}
Each diagonalizable $N\!\times\! N$ matrix is conjugate to a
linear combination of the projectors $P_k,~k=0,\dots N\! -\! 1$;
each $\su{N}$ matrix is conjugate to a linear combination of the
surfaces $X_k,\quad k=0,\dots N\! -\! 1$.
\end{proposition}
{\bf Proof}: Indeed, let $W$ be the diagonalizing matrix of a
given $N\!\times\! N$ matrix $M$, i. e.
\[ W^{-1}MW=\sum_{k=0}^{N-1}\lambda_k\mathbb{I}_k.\] Again, the matrix $V$ (\ref{V})
is the matrix which diagonalizes the projectors $P_k$ and so
\[ M=WV^{-1}\left(\sum_{k=0}^{N-1}\lambda_k P_k\right)VW^{-1}.\]
If we limit the sequences to those which satisfy
$\lambda_0+\ldots\lambda_{N-1}=0$, a similar argument shows that
each of the $\su{N}$ matrices is conjugate to a linear combination
of the surfaces $X_k$. \qed

   The matrices which represent spins are subalgebras of
$\su{N}$ ; more specifically, the usual quantum mechanics spins are
representations of $\su{2}$ in $\mathbb{C}^N$. The eigenvalues of
a matrix representing particles of spin $s$ are $-s,-s+1,\dots,s$.
Hence $s=(N-1)/2$. For such matrices we have the following.
\begin{proposition}
The matrices which represent spins are conjugate to
$-(i/2)\sum_{k=0}^{N-1}X_k$, i. e. every spin field is conjugate,
up to the factor $-i/2$, to a sum of all the surfaces $X_k,\,
k=0, \dots,N\! -\! 1$.
\end{proposition}
{\bf Proof}: Substituting $X_k$ from (\ref{Xk}) into
$-(i/2)\sum_{k=0}^{N-1}X_k$, expressing the unit matrix inherent
in (\ref{Xk}) as a sum of all the projectors
$\mathbb{I}=\sum_{k=0}^{N-1}P_k$, and collecting the terms at each
of $P_k$, we obtain
\be\fl
-\frac{i}{2}\sum\limits_{k=0}^{N-1}X_k=
\sum\limits_{k=0}^{N-1}\left(-\frac{1}{2}P_k-\sum\limits_{j=0}^{k-1}P_j+
\frac{1+2k}{2\,N}\sum\limits_{j=0}^{N-1}P_j\right)=\sum\limits_{k=0}^{N-1}\left(-\frac{N-1}{2}+k\right)P_k\,.
\ee
According to the above Proposition, the expression on the right-hand-side
is conjugate to any $\su{N}$ matrix whose eigenvalues are
$-(N-1)/2,-(N-1)/2+1,\dots,(N-1)/2$. \qed

Finally, note that the surface immersion functions $X_k$ span a
Cartan subalgebra of the $\su{N}$ algebra, i.e. not only do all
the $X_k$ commute with one another, but also any $N\!\times\! N$
matrix which commutes with all the $X_k,~k=0,...,N\! -\! 1$ is
their linear combination. The latter property stems directly from
the fact that being a Cartan subalgebra is invariant under
conjugation, while the $X_k$ are conjugate to the Abelian Lie
algebra of diagonal traceless $N\!\times\! N$ matrices.

\section{Conclusion}
In this paper, we considered soliton surfaces associated with
$\cpn$ sigma models. We show (Proposition \ref{prop1}) that the
GWFI for surfaces associated with $\cpn$ sigma models with finite
action defined on the Riemann sphere satisfies the E-L equations (\ref{EL})
Such surfaces are known to be conformally parameterized and to
have surface area proportional to the action functional.
Conversely, we have shown (Proposition \ref{prop2}) that the
immersion functions into $\su{N}$ in conformal coordinates that
are extremals of the area functional subject to any polynomial
identity satisfy the same E-L equations (\ref{EL}). This
discussion emphasized the importance of the eigenvalues of the
immersion functions in determining the class of solutions. In
particular, an immersion function $X$ is given by the GWFI for
solutions of the $\cpn$ sigma models with finite action defined on
the Riemann sphere if and only if it satisfies the E-L equations
(\ref{EL}) and has eigenvalues as indicated by (\ref{min}),
(\ref{minholo}) and (\ref{minantiholo}).

These quantized eigenvalues have a rich structure which was
elucidated in Section \ref{seceigen}. In that section, we
performed a systematic analysis of the eigenvalues, including the
action of the raising and lowering operators ($X_k\rightarrow
X_{k\pm 1}$) on the eigenvalues. We also described the symmetries
in the distribution as well as degeneracies of eigenvalues.
Finally, we discussed the distributions of eigenvalues for
different dimensions ($N$). Additionally, we show that any diagonalizable matrix is conjugate to a linear combination of the projectors and any $\su{N}$ matrix is conjugate to a linear combintation of the surfaces $X_k$; in particular matrices which represent spins can be written as a linear combination of the $X_k$'s.  Furthermore, the immersion functions $X_k$ form a Cartan subalgebra of $\su{N}$.

In the future, it would be interesting to consider a physical
interpretation for these eigenvalues in connection with spin
statistics as well as the connections with the orthogonal
polynomials associated with these distributions. These topics will
be explored in our future work.

 \ack This work was supported in
part by research grants from NSERC of Canada. PG thanks NSERC and
the Mathematical Physics Laboratory of the Centre de Recherches
Math\'ematiques, Universit\'e de Montr\'eal, for the financial
support provided for his visit to Montreal and for their warm
hospitality. SP acknowledges a postdoctoral ISM fellowship awarder
by the Mathematical Physics Laboratory of the Centre de Recherches
Math\'ematiques, Universit\'e de Montr\'eal.

\section*{References}

 \bibliography{suNbib}
 \bibliographystyle{jphysa}

\end{document}